\documentclass[12pt]{article}
\usepackage{amssymb,amsmath,epsfig}
\usepackage{graphicx}
\usepackage{cite}
\usepackage[utf8]{inputenc}
\usepackage{subfig}
 \allowdisplaybreaks
\begin{document}
\title{\bf Effects of $f (R, G)$ gravity on anisotropic charged compact objects}

\author{
  M. Ilyas\textsuperscript{1}\thanks{ilyas\_mia@yahoo.com},
  A. R. Athar\textsuperscript{2},
  F. Khan\textsuperscript{1},
  Asma Anfal\textsuperscript{1}\\
  \textsuperscript{1} Institute of Physics, Gomal University,\\
  Dera Ismail Khan, 29220, KP, Pakistan\\
  \textsuperscript{2} Institute of Physics, The Islamia University of Bahawalpur,\\
  Baghdad-ul-Jadeed Campus, Bahawalpur-63100, Pakistan}

\date{}

\maketitle
\begin{abstract}
The present study provides an in-depth analysis of the anisotropic matter distribution and various physical aspects of compact stars in the context of a $f(R,G)$-gravity framework. In order to gain an exhaustive understanding of these aspects, our study focuses on three particular compact stars: VELA X-1 (CS1), SAXJ1808.4-3658 (CS2), and 4U1820-30 (CS3). We conducted calculations on the relevant characteristics of these compact stars by employing three different models of $f(R,G)$-gravity. As a convenient approach, the $f(R,G)$-gravity is organized into two distinct components, which include $f_1(R)$ and $f_2(G)$. The $R$ dependent component is modeled similarly to the Hu-Sawicki approach, while for modeling the $G$ dependent component, we chose logarithmic and power law-like approaches and suggested three viable gravity models. Graphical methods are used to analyze the physical properties of the compact stars in the domain of suggested models of gravity.
\end{abstract}

\section{Introduction} Even though general relativity (GR) theory has been a great, well-known, and useful theory for the last hundred years, there have been many suggestions for major improvements in GR. One of the most promising areas of research is the unification of GR with quantum mechanics, which could lead to a theory of quantum gravity. Another area of interest is modifications to GR that could explain dark matter (DM) and dark energy (DE). Some of these modifications are: By incorporating an arbitrary function $f(R)$ instead of the Ricci scalar ($R$), similarly, by introducing a function of Gauss Bonnet (GB) invariant $f(G)$. The references \cite{p7r1,p7r6,p7r7,mia1,mia1a} include discussion of a number of such possible modifications in GR.
All such modified versions of GR are known as \emph{Modified Gravitational Theories} (MGTs). The published literature is a great source of information about MGTs \cite{p7r18,p7r19,p7r20,mia1d,mia1e,mia1f}.\\
The GB-gravity is considered a well-accomplished MGT that has been widely studied recently \cite{p7r22,p7r23,mia2}. It is constructed by introducing an arbitrary function $f(R,G)$ in the HE-action instead of $R$. Following is a mathematical description of the GB-term:
\begin{equation}\label{GBT}
G=R^2-4R_{\lambda\sigma}R^{\lambda\sigma}+R_{\lambda\sigma\eta\chi}R^{\lambda\sigma\eta\chi}.
\end{equation}
In Eq. (\ref{GBT}), $R_{\lambda\sigma}$ and $R_{\lambda\sigma\eta\chi}$ stand for the Ricci tensor and the Riemann tensor, respectively.\\
The $f(G)$-gravity is regarded as the most common type of $f(R,G)$-MGT, which has been studied a lot and can be used to model a wide range of cosmological solutions. As a scenario, it might play a role in the potential study of acceleration states and their gradual shift to deceleration states, the inflation period, and succeed throughout the investigations resulting from traversing the phantom divide line and planetary studies \cite{p7r26,p7r27}.
In a similar way, $f(G)$-gravity is an important part of studying the behavior of possible singularities in a finite amount of time, as well as the late-time behavior of the accelerating universe \cite{odintsov2023inflation,Odintsov:2022unp,Nojiri:2022xdo}.\\
The results of carefully measured observations of compact objects like neutron stars (NS), black holes (BH), and pulsars have changed the focus of scientific research from simple abstract explorations to the practical description of these real objects \cite{p7r36}.\\
In the context of MGTs and specifically $f(R,G)$-gravity, the most recent studies looked at the physical properties of different compact objects. After looking at their results, researchers came to the conclusion that all of the compact objects they were studying were stable and met the energy conditions (ECs) \cite{p7r39,sheikh2022electromagnetic,sharif2023role,gurian2022lower}. When making models of static stars, it is more practical to assume that the symmetry is spherical while having flexibility in choosing the composition of matter. The choice of perfect fluid as matter content is quite common in previous studies. On the other hand, anisotropic pressure and viscous fluids were also taken into account. It was concluded that anisotropy, as compared to a locally isotropic situation, could have an effect on the stability of the configuration. Also, many different kinds of physical processes and certain variations in density increase the local anisotropy \cite{herrera1997local}. Since then, the local anisotropic effects have also been worked out by using an equation of state (EoS) \cite{p7r40}. Therefore, it makes sense and is worthwhile to take pressure anisotropy into consideration, particularly when dealing with modified gravity models. Accounting for pressure anisotropy can help refine these models and provide a more accurate understanding of the universe's evolution. Modern studies of cosmology have given researchers a new way to look at astronomical objects from the point of view of phantom energy and DE \cite{debenedictis2008phantom}. The latest research suggests that various particular solutions have also been established using embedding class-1 \cite{maurya2017family}. Embedding class-1 solutions has provided a valuable tool for modeling and predicting the behavior of cosmological objects. The Karmarkar's limit proposal guarantees the possibility of embedding a spherically symmetric space-time with four dimensions into a flat space-time with five dimensions \cite{karmarkar1948gravitational}.
It is possible to relate both potentials of gravity using the geometric inference of the Karmarkar condition. This is helpful for getting a detailed analysis of the gravitational dynamics of the model in question, since one of the potentials needs specific metric functions and the other uses the Karmarkar condition. It is important to note that the Karmarkar limit with pressure isotropy means that the Schwarzchild interior solution leads to bounded matter emergence with pressure anisotropy that goes away \cite{abbas2021anisotropic}.\\
The scientists felt motivated to explore for realistic solutions to field equations as a result of their research into the  phenomena of particle physics that occur within the dense centers of various celestial objects. Incorporating the anisotropic distribution of pressure and assuming that these compact stars have a charge has given us a fascinating way to look at their physical properties \cite{biswas2021anisotropic,p7r41,p7r44,mia4,mia4a}. In the current study, a charged, spherically symmetric object is modeled by putting Schwarzschild coordinates into a 5-dimensional flat space-time that is also spherically symmetric. Additionally, the distribution of pressure is assumed to be anisotropic.\\
The main goal of this study is to use $f(R,G)$-gravity models to simulate the formation of charged, compact stars that make sense from a physical point of view. Here we are focused on investigating various properties of the compact stars related to their formation. For instance, we are examining how the density of charged matter is evolving, the significance of anisotropic pressure distribution, the role of the Tolman-Oppenheimer-Volkoff (TOV) equation,  different parameters of the EoS, and corresponding ECs.\\
This research article has been organized to include a subsequent section that discusses the framework of $f(R,G)$-gravity with respect to a charged anisotropic distribution of matter within the confines of static spherically symmetric geometry. We present an exploration of some feasible gravity models based on the $f(R,G)$-MGT in Section 3. Section 4 employs graphical representations to examine the physical properties and feasibility of several compact stars that have gained widespread acceptance. Lastly, the primary findings are thoroughly outlined in the concluding section.

\section{The $f(R,G)$ Framework}
The purpose of this section is to present an exhaustive discussion of $f(R,G)$-MGT, including its equations of motion. The modified HE-action of $f(R,G)$-MGT is typically expressed in the following manner:
\begin{equation}\label{act}
I = \int {d{x^4}\sqrt { - g} \left[{L_m+ L_e+ f(R,G)} \right]}.
\end{equation}
In the preceding Eq. (\ref{act}), $g=determinant(g_{\mu\nu})$, while $L_e$ and $L_m$ are the usual charge and matter lagrangian densities, respectively. In all of the computations that we are going to do for the present work, we are choosing to use relativistic units, which are defined as $8\pi G\equiv c \equiv 1$.\\
We are now using the variational principles method to vary action (Eq. (\ref{act})) with respect to the $g_{\mu\nu}$ to get the revised field equation as \cite{atazadeh2014energy}:
\begin{equation}\label{FEQS}
\begin{gathered}
{R_{\mu \nu }} - \frac{1}{2}{g_{\mu \nu }}R = {T_{\mu \nu }} + {E_{\mu \nu }} + 2R{\nabla _\mu }{\nabla _\nu }{f_G} - 2{g_{\mu \nu }}R{f_G} + {\nabla _\mu }{\nabla _\nu }{f_R} - {g_{\mu \nu }}{f_R}\\
 - 4R_\mu ^\alpha {\nabla _\alpha }{\nabla _\nu }{f_G} + 4{R_{\mu \nu }}{f_G} - 4R_\nu ^\alpha {\nabla _\alpha }{\nabla _\mu }{f_G} + 4{g_{\mu \nu }}{R^{\alpha \beta }}{\nabla _\alpha }{\nabla _\beta }{f_G}\\
 + 4{R_{\mu \alpha \beta \nu }}{\nabla ^\alpha }{\nabla ^\beta }{f_G} - \frac{1}{2}g\{ R{f_R} - f(R,G) + G{f_G}\}  - \{ {f_R} - 1\} {G_{\mu \nu }}.
\end{gathered}
\end{equation}
In the preceding Eq. (\ref{FEQS}), $f_G=\frac{\partial f}{\partial G}$, $f_R=\frac{\partial f}{\partial R}$, and
\begin{equation}
{T_{\mu \nu }} =({p_r} - {p_t}){V_\mu }{V_\nu } + (\rho  + p){U_\mu }{U_\nu } - {p_t}{g_{\mu \nu }}.
\end{equation}
Additionally, $V_\mu$ and $U_\mu$ refer to the four vector and four velocity of the fluid, correspondingly. In the co-moving coordinate system, the relationships $V_\mu V^\mu=-1$ and $U_\mu U^\mu=1$ are satisfied by the aforementioned quantities. Furthermore, in the above Eq. (\ref{FEQS}) ${G_{\mu \nu }} = {R_{\mu \nu }} - \frac{1}{2}{g_{\mu \nu }}R$ and $E_{\mu \nu }$ is given as,
\begin{equation}
 {E_{\mu \nu }} = \frac{1}{4\pi}\left[-g^{\gamma\delta}{{F_{\mu\gamma}}{F_{\gamma\delta}} + \frac{1}{4}g_{\mu\nu} {F_{\gamma\delta}}{F^{\gamma\delta}}} \right].
 \end{equation}
After we have set up the $f(R,G)$-gravity framework, we will discuss how anisotropic matter is distributed in this framework. The next section provides an exploration of possible avenues for further investigation into the behavior of anisotropic matter within the $f(R,G)$-gravity background.

\section{Anisotropic Matter Distribution}
Before we can study how matter is distributed, we need to define the corresponding metric. This metric is used to describe the dynamics of spherically symmetric geometry. The following is the standard expression for such a metric, which in our case is the Krori-Barua (KB) metric\cite{krori1975singularity,biswas2019strange,sheikh2023evolution}:
\begin{equation}\label{KBmtr}
d{s^2} = e^{a(r)}d{t^2} -e^{b(r)}d{r^2}-{r^2}( {d{\theta ^2} + {{r^2}{\sin }^2}\theta d{\phi ^2}}).
\end{equation}
The above-mentioned metric includes two free variables, $a(r)$ and $b(r)$. Now that we consider that the conditions do not involve any singularities at $r\rightarrow 0$, the field equation drives us to hold the more convenient arrangement for $a(r)$ and $b(r)$ \cite{krori1975singularity}:
\begin{equation}\label{ar}
  a(r)=B r^2 +C,
\end{equation}
\begin{equation}\label{br}
  b(r)=A r^2.
\end{equation}
The symbols $A$, $B$, and $C$ denote arbitrary constants. We estimate the numerical values of these constants by taking into account various physical conditions.\\
Further we solve the field equation (\ref{FEQS}) to obtain the following:
\begin{equation}\label{rauf1}
\begin{gathered}
\rho  + {E^2} = \frac{1}{{2{r^2}}}{{\rm{e}}^{ - 2b}}({{\rm{e}}^{2b}}{r^2}G{f_G} - {{\rm{e}}^{2b}}{r^2}f(R,G) + (2{{\rm{e}}^b} + {{\rm{e}}^b}{r^2}R + 2rb' - 2){{\rm{e}}^b}{f_R}\\
 - 4{{\rm{e}}^b}r{f_R}^\prime  + 12b'{f_G}^\prime  - 4{{\rm{e}}^b}b'{f_G}^\prime  + {{\rm{e}}^b}{r^2}b'{f_R}^\prime  - 2{{\rm{e}}^b}{r^2}{f_R}^{\prime \prime } - 8{f_G}^{\prime \prime } + 8{{\rm{e}}^b}{f_G}^{\prime \prime }),
\end{gathered}
\end{equation}
\begin{equation}\label{rauf2}
\begin{gathered}
{p_r} - {E^2} = \frac{1}{{2{r^2}}}{{\rm{e}}^{ - 2b}}{\rm{[}}{r^2}{{\rm{e}}^{2b}}f(R,G) - {{\rm{e}}^{2b}}{r^2}G{f_G} - {{\rm{e}}^b}{f_R}\{ 2{{\rm{e}}^b} - 2\\
 + {{\rm{e}}^b}{r^2}R - 2ra'\}  + 4{{\rm{e}}^b}r{f_R}^\prime  + {{\rm{e}}^b}{r^2}a'{f_R}^\prime  + 12a'{f_G}^\prime  - 4{{\rm{e}}^b}a'{f_G}^\prime ],
\end{gathered}
\end{equation}
and
\begin{equation}\label{rauf3}
\begin{gathered}
{p_t} + {E^2} = \frac{1}{{4r}}{{\rm{e}}^{ - 2b}}[2r{{\rm{e}}^{2b}}f(R,G) - 2{{\rm{e}}^{2b}}r{f_R}R + 2{{\rm{e}}^b}{f_R}a' - 2{{\rm{e}}^{2b}}rG{f_G}\\
 - {{\rm{e}}^b}r{f_R}a'b' + {{\rm{e}}^b}r{f_R}{{a'}^2} + 4{{\rm{e}}^b}{f_R}^\prime  - 2{{\rm{e}}^b}{f_R}b' + 4{{a'}^2}{f_G}^\prime  - 12a'b'{f_G}^\prime \\
 + 2{{\rm{e}}^b}r{f_R}{a^{\prime \prime }} + 2{{\rm{e}}^b}ra'{f_R}^\prime  - 2{{\rm{e}}^b}rb'{f_R}^\prime  + 4{{\rm{e}}^b}r{f_R}^{\prime \prime } + 8{f_G}^\prime {a^{\prime \prime }} + 8a'{f_G}^{\prime \prime }],
\end{gathered}
\end{equation}
where $\rho$, $p_r$, and $p_t$ are energy density, radial pressure, and transverse pressure, respectively, while it can be easily shown that the electric field intensity can be written in term of electric charge e.g. $E^2=\frac{Q^2}{8\pi r^4}$ and the Ricci scalar $R$ is
\begin{equation}
R = \frac{{{{\rm{e}}^{ - b}}}}{{2{r^2}}}[{r^2}{a'^2} - 4rb' + ra'\{ 4 - rb'\}  - 4{{\rm{e}}^b} + 2{r^2}{a^{\prime \prime }} + 4],
\end{equation}
and
\begin{equation}
G = \frac{{2{{\rm{e}}^{ - 2b}}}}{{{r^2}}}\left[ { + \{ {{\rm{e}}^b} - 3\} a'b' - \{ {{\rm{e}}^b} - 1\} {{a'}^2} - 2\{ {{\rm{e}}^b} - 1\} {a^{\prime \prime }}} \right].
\end{equation}
Note that the prime symbol in the equations above and below stands for a derivative in terms of the radial coordinates. Moreover, we use Eqs. (\ref{ar})-(\ref{br}) in Eqs. (\ref{rauf1}, \ref{rauf2}, \ref{rauf3}) and obtain the following simplified expressions:
\begin{equation}\label{ro}
\begin{gathered}
\rho  = \frac{{{{\rm{e}}^{ - 2A{r^2}}}}}{{8\pi {r^4}}}\left[ {{{\rm{e}}^{2A{r^2}}}} \right.( - 4\pi {r^4}\left( {f - {f_G}G} \right) - {Q^2} + 4\pi {r^2}{f_R}\\
(2 + {r^2}R)) + 32\pi {r^2}(3Ar{f_G}^\prime  - {f_G}^{\prime \prime }) + 8{{\rm{e}}^{A{r^2}}}\pi {r^2}\\
((2A{r^2} - 1){f_R} + 4{f_G}^{\prime \prime } + r( - 4A{f_G}^\prime  + ( - 2 + A{r^2}){f_R}^\prime  - r\left. {{f_R}^{\prime \prime }))} \right]
\end{gathered}
\end{equation}
\begin{equation}\label{pr}
\begin{gathered}
{p_r} = \frac{{{{\rm{e}}^{ - 2A{r^2}}}}}{{8\pi {r^4}}}\left[ {{{\rm{e}}^{2A{r^2}}}} \right.(4\pi {r^4}\left( {f - {f_G}G} \right) + {Q^2}\\
 - 4\pi {r^2}{f_R}(2 + {r^2}R)) + 96B\pi {r^3}{f_G}^\prime  + 8{{\rm{e}}^{A{r^2}}}\\
\pi {r^2}({f_R} + 2B{r^2}{f_R} - 4Br{f_G}^\prime  + r(2 + B{r^2})\left. {{f_R}^\prime )} \right]
\end{gathered}
\end{equation}
\begin{equation}\label{pt}
\begin{gathered}
{p_t} = \frac{{{{\rm{e}}^{ - 2A{r^2}}}}}{{8\pi {r^4}}}\left[ {{\rm{4}}{{\rm{e}}^{2A{r^2}}}} \right.f\pi {r^4} - 4{{\rm{e}}^{2A{r^2}}}\pi {r^4}{f_G}G - {{\rm{e}}^{2A{r^2}}}{Q^2}\\
 - 4{{\rm{e}}^{A{r^2}}}\pi {r^4}{f_R}(2(A + AB{r^2} - B(B{r^2} + 2)) + {{\rm{e}}^{A{r^2}}}R)\\
 + 32B\pi {r^3}{f_G}^\prime  - 96AB\pi {r^5}{f_G}^\prime  + 32{B^2}\pi {r^5}{f_G}^\prime  + 8{{\rm{e}}^{A{r^2}}}\pi {r^3}{f_R}^\prime \\
 - 8A{{\rm{e}}^{A{r^2}}}\pi {r^5}{f_R}^\prime  + 8B{{\rm{e}}^{A{r^2}}}\pi {r^5}{f_R}^\prime  + 32B\pi {r^4}{f_G}^{\prime \prime } + 8{{\rm{e}}^{A{r^2}}}\pi {r^4}\left. {{f_R}^{\prime \prime })} \right]
\end{gathered}
\end{equation}
The contribution of charge is now visible in the above-mentioned expressions of $\rho$, $p_r$ and $p_t$ (Eqs. \ref{ro} - \ref{pt}). In addition, these charge characteristics will be taken into consideration.\\
Now, we examine a fundamental representation of the EoS that deals with the strange matter, which is as follows:
\begin{equation}\label{EoS}
{p_r} = \frac{1}{3}\left( {\rho  - 4{B_g}} \right).
\end{equation}
The symbol $B_g$ in the above expression (\ref{EoS}) represents the bag constant. By using the Eq. (\ref{EoS}), we get the overall expression of the charge, as shown below:
\begin{equation}
\begin{gathered}
Q = {{\rm{e}}^{ - A{r^2}}}\sqrt {2\pi } r\left[ { - 2{{\rm{e}}^{2A{r^2}}}} \right.({r^2}(2{B_g} + f - {f_G}G) - {f_R}(2 + {r^2}R))\\
 + 12(A - 3B)r{f_G}^\prime  - 4{f_G}^{\prime \prime } + {{\rm{e}}^{A{r^2}}}(2( - 2 + (A - 3B){r^2}){f_R}\\
 + 4{f_G}^{\prime \prime } + r( - 4(A - 3B){f_G}^\prime  + ( - 8 + (A - 3B){r^2}){f_R}^\prime  - rf{\left. {{{_R}^{\prime \prime }}))} \right]^{\frac{1}{2}}}.
\end{gathered}
\end{equation}
It can be seen that the above general expression for the charge has two arbitrary constants, $A$ and $B$. By matching an interior metric to an exterior metric, one can get the simplified expressions for these constants. In the next section, we use matching conditions to obtain expressions of these constants.
\section{Geometric Matching}
Considering the internal and external geometries of the compact celestial object, it is obvious that the internal metric remains unchanged at the boundary of the object. This shows that the components of the metric do not change at the star's boundary. One can make a number of feasible choices to interpret the matching conditions. For instance, one can use a general, spherically symmetric space-time and then analyze the vacuum at the exterior of that space-time through specific conditions at the boundary \cite{yazadjiev2014non}. The way we are doing this in our current study is to set up a hypersurface that separates the interior and exterior of the system. We are using Krori-Barua (KB) space-time to represent the interior geometry of this surface, while Reissner-Nordstr\"{o}m space-time represents the exterior geometry of this surface. The choice of KB-metric is very plausible because it does not incorporate any singularity. The prior studies presented numerous thought-provoking findings through the use of Reissner-Nordstr\"{o}m solutions. The exterior geometry is expressed by the following:
\begin{equation}\label{ExMtr}
d{s^2} = \left[ {1 - \frac{{2M}}{r}+\frac{{{Q^2}}}{{{r^2}}}} \right]d{t^2} - \frac{d{r^2}}{{\left[ {1 - \frac{{2M}}{r}+\frac{{{Q^2}}}{{{r^2}}}} \right]}} -{r^2}\left[ {d{\theta ^2} + {\mathop{\rm \sin}\nolimits} {\theta ^2}d{\varphi ^2}} \right].
\end{equation}
In the situation above (\ref{ExMtr}), $M$, $r$, and $Q$ stand for the mass, the radius, and the charge, respectively. When it comes to the distribution of charged fluids, the equation Eq. (\ref{KBmtr}) and the equation Eq. (\ref{ExMtr}) connect perfectly. We then match the two geometries by using $r=R$ and $M(R)=M$ as matching conditions. Thus matching results in the following expressions for the constants $A$, $B$, and $C$:
\begin{equation}
A = \frac{{ - 1}}{{{R^2}}}\ln \left[ {1 - \frac{{2M}}{R}+\frac{{{Q^2}}}{{{R^2}}}} \right],
\end{equation}
\begin{equation}
B = \frac{1}{{{R^2}}}\left[ {\frac{M}{R} + \frac{{{Q^2}}}{{{R^2}}}} \right]{\left[ {1 - \frac{{2M}}{R} + \frac{{{Q^2}}}{{{R^2}}}} \right]^{ - 1}},
\end{equation}
\begin{equation}
C = \ln \left[ {1 - \frac{{2M}}{R} + \frac{{{Q^2}}}{{{R^2}}}} \right] - \left[ {\frac{M}{R} - \frac{{{Q^2}}}{{{R^2}}}} \right]{\left[ {1 - \frac{{2M}}{R} + \frac{{{Q^2}}}{{{R^2}}}} \right]^{ - 1}}.
\end{equation}
It is important to compute the numerical values of the above-expressed constants. Then, we figure out $A$ and $B$ by using the estimated masses ($M$) and radii ($R$) of the three compact stars we are studying. The numerical values of $A$ and $B$ are given in Table: (\ref{table:1}).
\begin{table}[h!]
\centering
\begin{tabular}{c c c c c c c}
 \hline
Compact Stars  & Radius$(km)$ & Mass  & $\mu_M=\frac{M}{R}$ & $\mu_C=\frac{Q^2}{R^2}$ & $A$ &$B$ \\ [0.5ex]
\hline\hline\
Vela X - 1 & $9.56$ \cite{gangopadhyay2013strange}& $1.77M_{\odot}$ \cite{gangopadhyay2013strange}& $0.273091$ & $0.0133624$ &$0.00832706 $ &$0.00608302 $ \\ [1ex]
\hline\
SAXJ1808.4-3658 & $7.07$ \cite{gangopadhyay2013strange} & $1.435M_{\odot}$ \cite{gangopadhyay2013strange}& $0.299$ & $0.0266898$ & $0.0169456$ &$0.0127081$\\ [1ex]
\hline
4U1820-30 & $10$ \cite{gangopadhyay2013strange}& $2.25M_{\odot}$ \cite{gangopadhyay2013strange}& $0.332$ & $0.0133208$ & $0.00760739 $ &$0.00555676$\\ [1ex]
\hline
\end{tabular}
\caption{The masses, radii, compactness, and constants of compact stars.}
\label{table:1}
\end{table}
Now that we have set up almost all of the requirements for our study of the physical properties of compact stars, we need to look at some models that work in the $f(R,G)$-gravity domain. In the upcoming section, we shall present a detailed analysis of these models.
\begin{table}[h!]
\centering
\begin{tabular}{c c c c}
 \hline
Compact Stars  & Model-I & Model-II & Model-III \\ [0.5ex]
\hline\hline\
$B_g$ for CS-I & 0.0033638 & 0.0033638 & 0.00336385 \\ [1ex]
\hline\
$B_g$ for CS-II & 0.00634729 & 0.0063473 & 0.00634825\\ [1ex]
\hline
$B_g$ for CS-III & 0.00307407 & 0.00307407 & 0.0030741\\ [1ex]
\hline
\end{tabular}
\caption{Estimates of the $B_g$ for considered compact stars using suggested gravity models. }
\label{table:2}
\end{table}

\section{The Models of $f(R,G)$-Gravity}
In this section, we are considering three feasible models using the framework of $f(R,G)$-gravity. These models are based on space-time geometry, which is spherically symmetric. Using these models, we examine various important physical characteristics of the compact stars under investigation. Mainly, we shall focus on examining properties such as mass, the evolution of energy density and pressure, the ECs, the role of the TOV-equation, the stability analysis, the EoS parameter, the anisotropic measurement, and the impact of the electric field and charge. The results obtained from this analysis will provide a better understanding of the behavior of compact stars and their properties. Furthermore, the findings may have significant implications for astrophysical observations and the development of future theoretical models.  In order to avoid tedious arrangement, we split the $f(R,G)$-gravity as under:
$$f(R,G)=f_1(R)+f_2(G)$$
The first part $f_1(R)$ of the above division is modeled after the Hu and Sawicki approaches. Following is the mathematical expression for $f_1(R)$:
$$f_1(R) =  R + \alpha \gamma \left[ \left(1 + \frac{R^2}{\gamma^2}\right)^{-q} -1\right].$$
The symbols $\alpha$, $\gamma$, and $q$ in the above mentioned expression of $f_1(R)$  are free parameters of the model. Choosing $f_1(R)$ in this way makes sure that the cosmological solutions that come out of it will be feasible, especially for the matter-dominant epoch that shows a late-time acceleration \cite{martinelli2009cosmological}. Also, the approach that Hu and Sawicki came up with has the same kind of acceleration as that exhibited by the $\Lambda$CDM at high red-shifts. In all three models, we come up with the other part of the $f(R,G)$ division, $f_2(G)$, by making different plausible choices.

\subsection{Model-I}
For Model-I, we suggest that the $f_2(G)$ be chosen in a logarithmic way as follows: $f_2(G)=\beta G^\lambda+\gamma G\log(G)$, with constants $\beta$ and $\gamma$ \cite{schmidt2011gauss}. The full $f(R,G)$-gravity model takes the following form:
\begin{equation}\label{model1}
f(R,G) =  R + \alpha \gamma \left[ \left(1 + \frac{R^2}{\gamma^2}\right)^{-q} -1\right] + \beta {G^\lambda } + \gamma G\log \left[ G \right].
\end{equation}
In the preceding expression (\ref{model1}) of the Model-I, $G$ represents that GB-curvature. The full $f(R,G)$-gravity model takes into account the scale-invariant field equation represented by $\lambda$. It is important to mention here that the logarithmic term has no dimensions. The choice of a logarithmic-like model for $f(G)$ makes the full model a simpler and more concise one. The model is more applicable and consistent with the available data.

\subsection{Model-II}
For Model-I, we suggest that the $f_2(G)$ be chosen in a power law model way as follows: $f_2(G)=\beta G^n[1+\gamma G^m]$, with real constants $n$ and $m$ \cite{p7r47}. These constants serve as model parameters. The full $f(R,G)$-gravity model takes the following form:
\begin{equation}\label{model2}
f(R,G) = R + \alpha \gamma \left[ \left(1 + \frac{R^2}{\gamma^2}\right)^{-q} -1\right] +\beta {G^n}\left[ {1 + \gamma {G^m}} \right].
\end{equation}
In the preceding expression (\ref{model2}) of the Model-II, it would be interesting to find conditions for $\gamma$, $\beta$, $m$, and $n$, that make all singularities disappear. The most important thing to note here is that we do not assume an insignificant scenario, which is: $n=m$, but consider $n>0$. So, the suggested model is more likely to work and fits with the literature and data that have already been presented.

\subsection{Model-III}
For Model-III, we suggest that the $f_2(G)$ be chosen again in a power law model way as follows: $f_2(G) = \frac{(a_1G^n+b_1)}{a_2G^n+b_2}$, with constant parameters $a_1$, $a_2$, $b_1$, and $b_2$ \cite{p7r47}. The full $f(R,G)$-gravity model takes the following form:
\begin{equation}\label{model3}
f(R,G) =  R + \alpha \gamma \left[ \left(1 + \frac{R^2}{\gamma^2}\right)^{-q} -1\right] +\left[\frac{\left({a_1}{G^n} + {b_1}\right)}{{\left( {{a_2}{G^n} + {b_2}} \right)}}\right].
\end{equation}
When $G$ is large, the expression tends to converge to a fixed value, which makes it unfeasible to find singularities. The $\Lambda$CDM model is consistent with the $R+constant$ because it doesn't involve any singularities. The singularities may still emerge for $G\rightarrow0^{-}$ when $n>1$. So, the suggested model is more likely to work and fits with the literature and data that have already been presented.

We use these suggested gravity models to study different physical properties of the three compact stars. Also, the physical analysis of the models helps check the feasibility of these models. The next section presents a comprehensive such analysis.

\section{Physical Analysis of the Models}
Here, we use the gravity models from the previous section to look in depth at some important physical properties of the internal solutions that are discussed in more detail in the subsections that follow.

\subsection{The Evolution of Energy Density and Pressure}
\begin{figure}[h!]
\centering
\epsfig{file=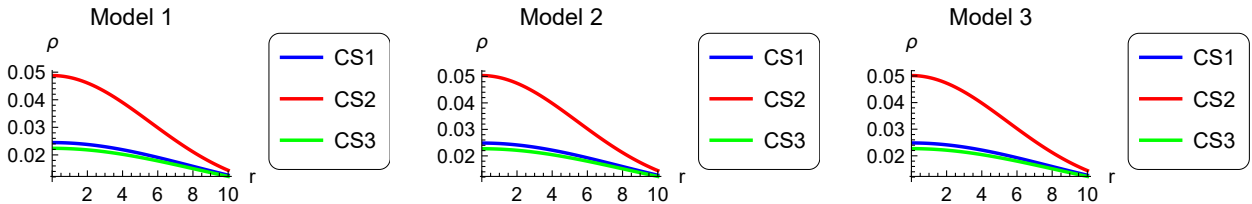,width=1\linewidth}
\caption{Density evolution of three strange star candidates using suggested models}\label{roc}
\end{figure}

\begin{figure}[h!]
\centering
\epsfig{file=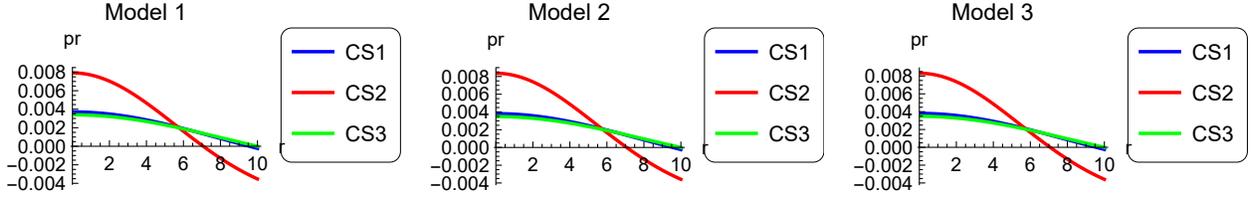,width=1\linewidth}
\caption{Radial pressure evolution of three strange star candidates using suggested models}\label{prc}
\end{figure}

\begin{figure}[h!]
\centering
\epsfig{file=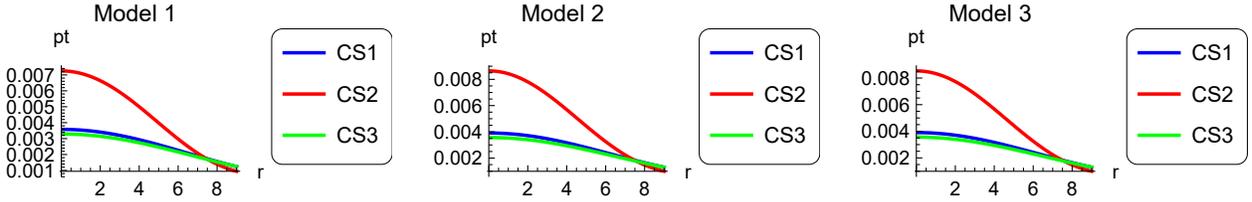,width=1\linewidth}
\caption{Transverse pressure evolution of three strange star candidates using suggested models}\label{ptc}
\end{figure}

\begin{figure}[h!]
\centering
\epsfig{file=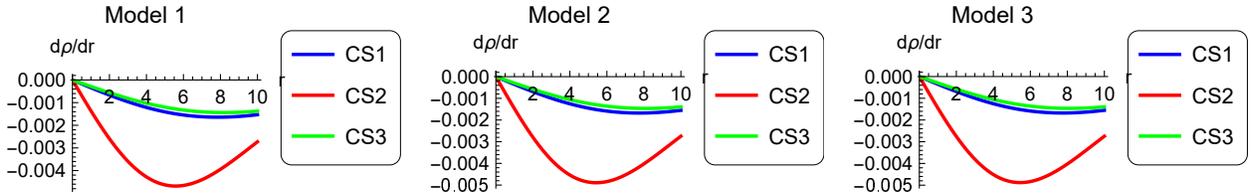,width=1\linewidth}
\caption{The $d\rho/dr$ variation as a function of $r$ for the three strange star candidates using suggested models}\label{dro}
\end{figure}

\begin{figure}[h!]
\centering
\epsfig{file=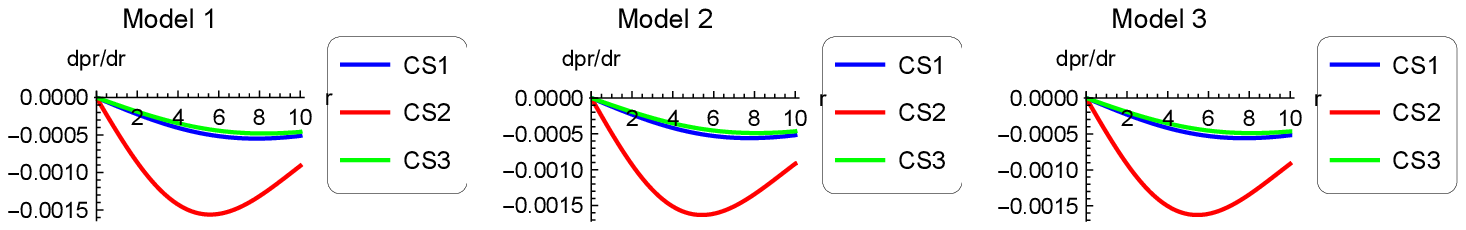,width=1\linewidth}
\caption{The $dp_r/dr$ variation as a function of $r$ for the three strange star candidates using suggested models}\label{dpr}
\end{figure}

\begin{figure}[h!]
\centering
\epsfig{file=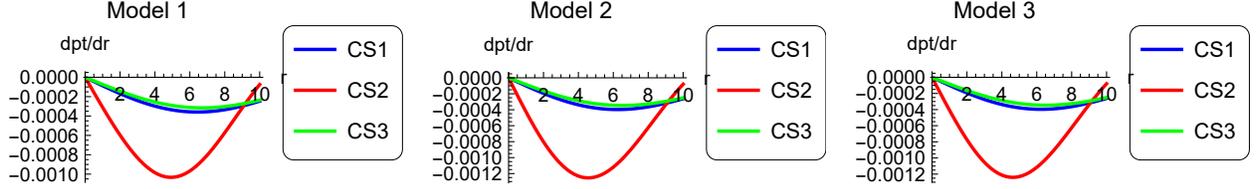,width=1\linewidth}
\caption{The $dp_t/dr$ variation as a function of $r$ for the three strange star candidates using suggested models}\label{dpt}
\end{figure}

When we look at the plots in Fig. (\ref{roc}) of the compact stars under investigation, the plots show that the density ($\rho$) becomes maximum with approaching the core($r \rightarrow 0$). Mathematically, it can be concluded that the density function is inversely related to $r$. As $r$ increases (away from the center), the density decreases. This variation of $\rho$ with respect to $r$ shows that the core of the compact star has a high compactness ($\mu$) value. This shows that our suggested models are very realistic and help us gain a better understanding of compact stars, especially in the case of the core's outer domain.\\
Following the above, see Figs. (\ref{prc})-(\ref{ptc}), which show the behavior of anisotropic pressures. In these plots, it is demonstrated how the radial pressure ($p_r$) and the transverse pressure ($p_t$) change with respect to $r$.
As part of our analysis, we have shown the plots of the radial derivatives of the density ($\frac{d\rho}{{dr}}$), the radial pressure ($\frac{dp_r}{{dr}}$), and the transverse pressure ($\frac{dp_t}{{dr}}$) in Figs. (\ref{dro})-(\ref{dpt}), respectively. Our analysis shows that all three of our suggested models, in the case of all three compact stars, have the following results: $\frac{d\rho}{{dr}}<0$, $\frac{dp_r}{{dr}}<0$ and $\frac{dp_t}{{dr}}<0$. We also figure out that at the center ($r = 0$) of these compact stars, the quantities $\frac{d\rho}{{dr}}$ and $\frac{dp_r}{{dr}}$ result in the following:
$$\frac{d\rho}{{dr}}=0$$
$$\frac{dp_r}{{dr}}=0$$
The above-mentioned results are very likely, as the density function is naturally decreasing. The maximum values of $\rho$ comes from very small values of $r$. This maximum density is referred to as the core density ($\rho(0)=\rho_c$).

\subsection{The Energy Conditions}
\begin{figure}[h!]
\centering
\epsfig{file=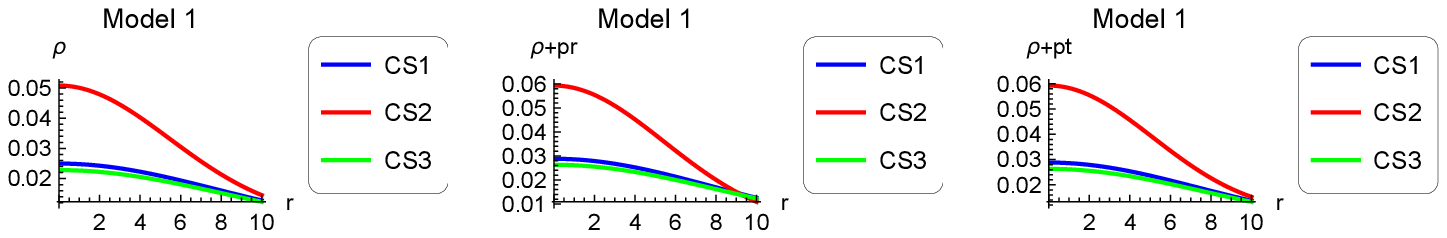,width=1\linewidth}
\epsfig{file=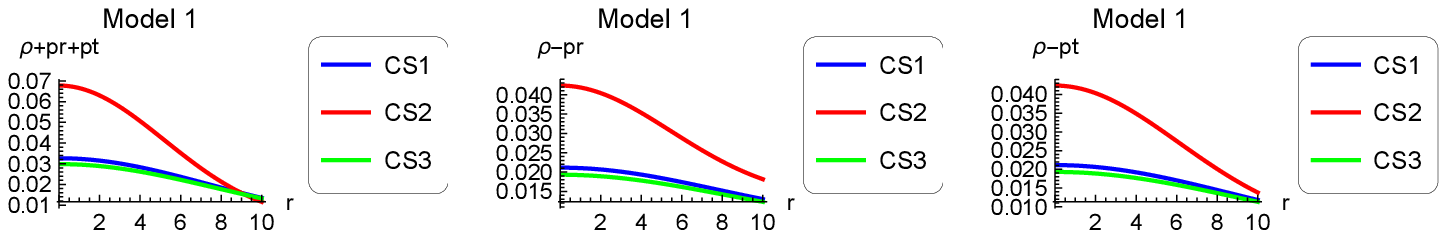,width=1\linewidth}
\caption{ECs in case of Model-I}\label{energy1}
\end{figure}

\begin{figure}[h!]
\centering
\epsfig{file=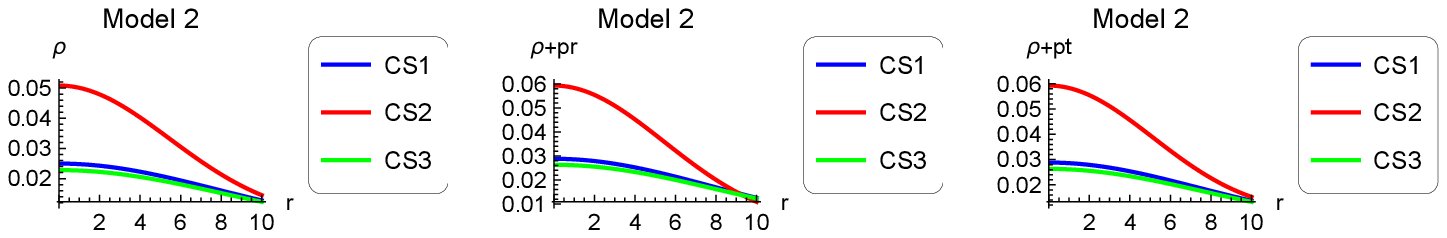,width=1\linewidth}
\epsfig{file=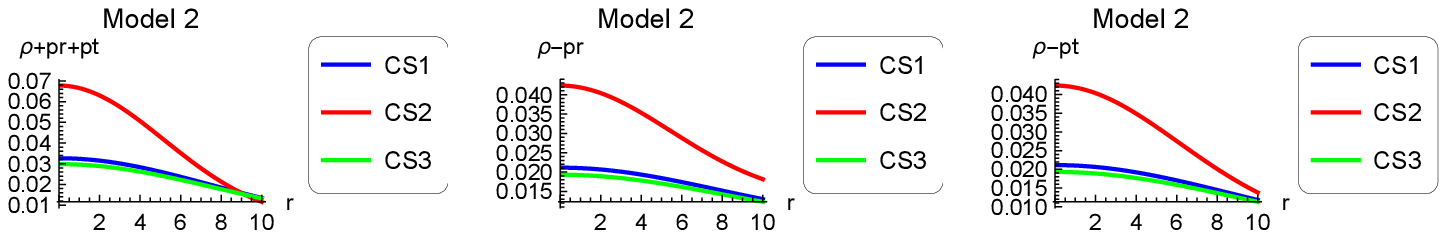,width=1\linewidth}
\caption{ECs in case of Model-II}\label{energy2}
\end{figure}

\begin{figure}[h!]
\centering
\epsfig{file=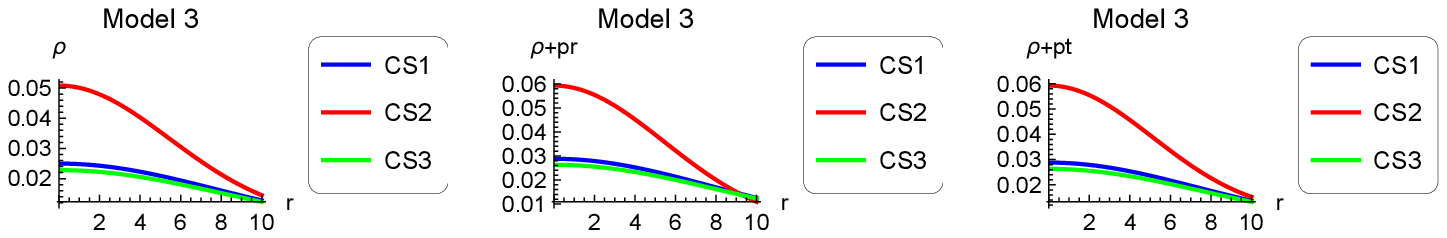,width=1\linewidth}
\epsfig{file=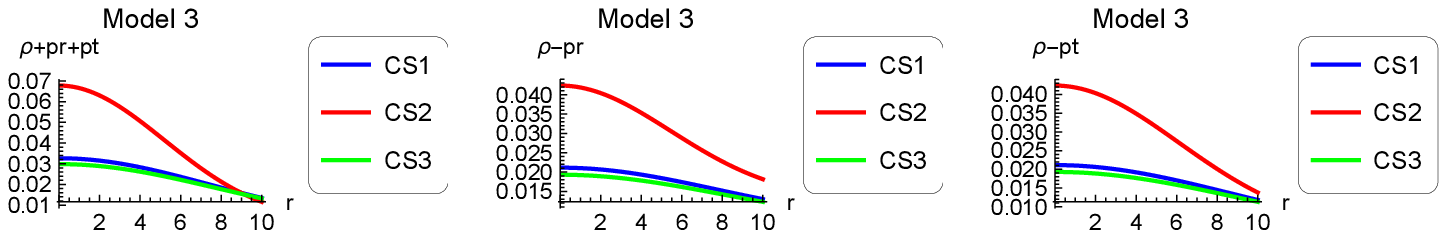,width=1\linewidth}
\caption{ECs in case of Model-III}\label{energy3}
\end{figure}

Researchers examine ECs regarding a variety of various problems associated with the astrophysical domain of study. The contribution of ECs to a variety of stimulating outcomes \cite{p7r26,atazadeh2014energy,balart2014regular,lobo2011late,bertolami2009energy} is evidence that their role in cosmology has been interesting. Deriving a thorough arrangement of ECs involved using Raychaudhuri's well-known equation for expansion. Based on the ECs, one can infer that for positive density values, gravity shows an attractive nature but is unable to persist at velocities exceeding $c$ (light's speed). There are four fundamental ECs, which can be listed as follows:
\begin{enumerate}
  \item Null ECS (NEC) ${\rho + p_r} \ge 0$,  ${\rho + p_t} \ge 0$.
  \item Weak ECs (WEC) $\rho \ge 0$, ${\rho + p_r} \ge 0$,    ${\rho + p_t} \ge 0$.
  \item Strong ECs(SEC) ${\rho + p_r} \ge 0$,  ${\rho + p_t}\ge 0$, ${\rho + p_r + 2p_t}\ge0$.
  \item Dominant ECs (DEC)$\rho \ge |{p_r}|$, $\rho \ge |{p_t}|$.
\end{enumerate}
Out of these ECs, the NEC and SEC are widely investigated, and results are thoroughly furnished in previous research. The behavior of all of these ECs is thoroughly investigated in our current study. We found that these ECs are valid for the compact stars under investigation in our suggested gravity models. We use graphical methods to establish the validity of ECs. The behavior of these ECs is shown in Figs. (\ref{energy1})-(\ref{energy3}) by plotting corresponding quantities. It is important to mention that the above forms of ECs involve $\rho$, $p_r$, and $p_t$ that incorporate electric charge as well.

\subsection{Role of the TOV Equation}
\begin{figure}[h!]
\centering
\epsfig{file=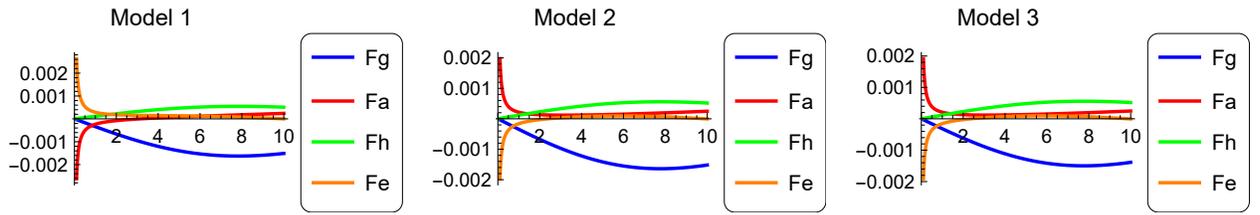,width=1\linewidth}
\caption{Plots of $F_g$, $F_h$, and $F_a$ with respect to $r (km)$}\label{eqb}
\end{figure}

The equation for the TOV scheme is expressed in a generalized form as follows:
\begin{equation}\label{tov}
\frac{{d{p_r}}}{{dr}} + \frac{{\nu'(\rho  + {p_r})}}{2} + \frac{{2({p_r} - {p_t})}}{r} + \frac{{\sigma Q}}{{{r^2}}}{e^{\lambda /2}} = 0.
\end{equation}
In the above expression (\ref{tov}), the symbol $\sigma$ represents the charge density. It is convenient to reexpress the above expression (\ref{tov}) by introducing the three forces, which are: gravitational ($F_g$), hydrostatic ($F_h$), electric ($F_e$), and anisotropic ($F_a$). All these forces can be related using the following simple mathematical expression:
\begin{equation}\label{retov}
F_g + F_h + F_a+F_e= 0.
\end{equation}
We compare the expressions (\ref{tov}) and (\ref{retov}) to obtain the following:
$$F_g=-B r (\rho+p_r),  F_h=-\frac{{d{p_r}}}{{dr}},   F_a= \frac{{2({p_r} - {p_t})}}{r}, F_e= \frac{{\sigma Q}}{{{r^2}}}{e^{\lambda /2}}$$
We extend our analysis using graphical methods and by employing forms of force obtained through comparison. We analyze the behavior of these forces for the compact stars under investigation using our suggested gravity models. The plots given in Fig. (\ref{eqb}) show the behavior of these forces with respect to varying radial coordinate $r$.

\subsection{The Aspect of Stability}
In this subsection, we examine the stability of our suggested gravity models in the case of three compact stars under investigation. To check the stability, we define the radial and transverse speeds in the following manner:
$$\frac{{d{p_r}}}{{d\rho }} = v_{sr}^2$$
and
$$\frac{{d{p_t}}}{{d\rho }} = v_{st}^2$$
In order to have a stable configuration of compact stars in the case of suggested gravity models, it is mandatory that both speeds satisfy the following conditions:
$$0 \le v_{sr}^2 \le 1,$$ and $$0 \le v_{st}^2 \le 1.$$
\begin{figure}[h!]
\centering
\epsfig{file=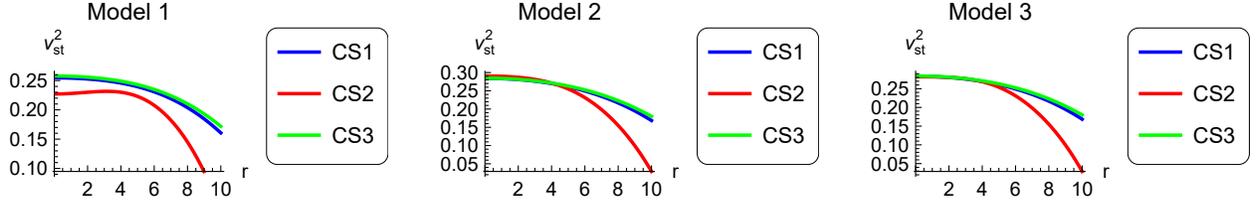,width=1\linewidth}
\caption{Squared transverse speed as a function of $r(km)$ for three strange star candidates}\label{vst}
\end{figure}

Furthermore, we plot these speeds to check how they evolve with respect to radial coordinates for considered compact stars and in the domain of suggested gravity models. The plots shown in Fig. (\ref{vst}) establish evolution of these speeds. As $v_{sr}^2\sim 1/3$, it can be seen from the Fig. (\ref{vst}) that the evolution of the $v_{sr}$ and $v_{st}\sim 1/3$ sound speeds for all three type of compact stars candidate are within the bounds of stability as discussed. Herrera proposed that it is essential for a stable configuration to meet the following requirements \cite{p7r88}:
$$v_{sr}>v_{st}.$$
So, keeping in mind this important requirement, we further plotted this condition to check the validity of our suggested models in the case of three compact star candidates. These plot are shown in Fig. (\ref{vstmvsr}). We obtain the following result from plots:
\begin{figure}[h!]
\centering
\epsfig{file=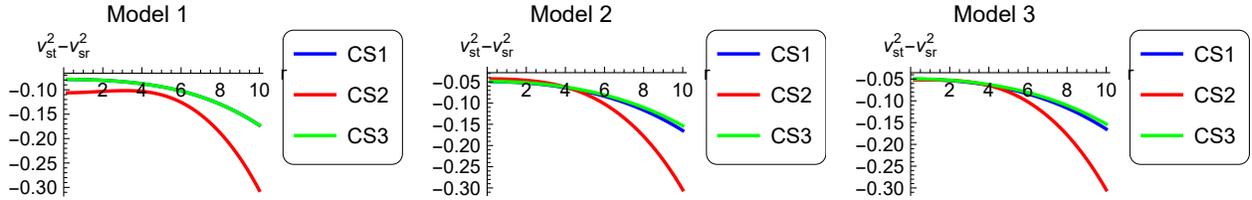,width=1\linewidth}
\caption{Difference of squared transverse and radial speeds with respect to $r(km)$ for three strange star candidates}\label{vstmvsr}
\end{figure}
$$0< |v_{st}^2-v_{sr}^2|<1.$$
The above result confirms that the compact stars under investigation are stable in the domain of our suggested gravity models.

\subsection{The EoS Parameter}
\begin{figure}[h!]
\centering
\epsfig{file=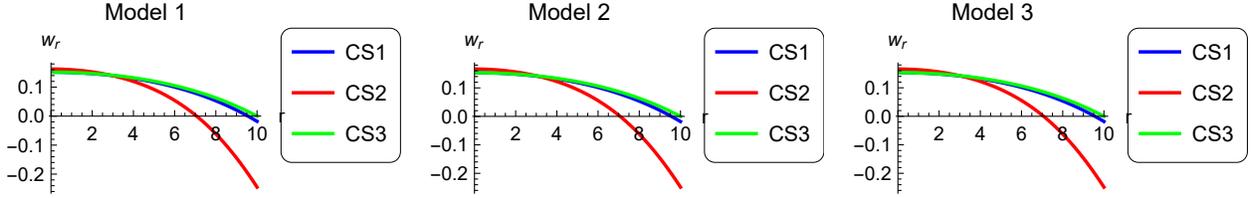,width=1\linewidth}
\caption{The plots of $w_r$ as a function of $r$}\label{wr}
\end{figure}

\begin{figure}[h!]
\centering
\epsfig{file=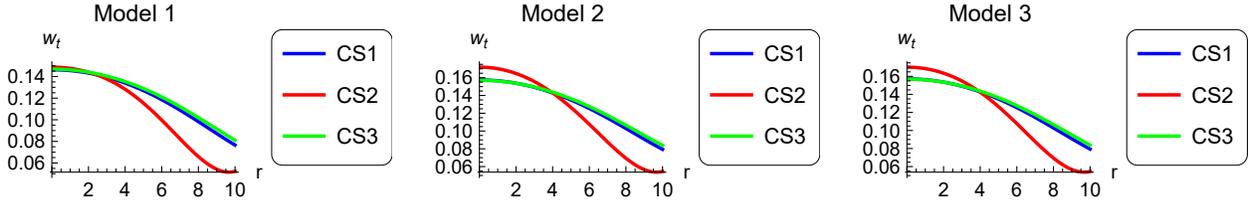,width=1\linewidth}
\caption{The plots of $w_t$ as a function of $r$}\label{wt}
\end{figure}

The exhaustive literature review shows that it is very likely that there are a number of frameworks in cosmology that include various parameters of the EoS. For instance, in Staykov \emph{et al}. \cite{yazadjiev2014non}, they investigated different configurations of rotating neutron stars by employing an EoS parameter of the strange object with two additional hadronic parameters. The survey further establishes that, using quadratic EoS parameters, various physical properties of compact objects are investigated \cite{malaver2014strange,ngubelanga2015compact}. For the current investigation, we consider that EoS takes the linear forms when dealing with anisotropic matter content, as described
$$p_r=w_r \rho$$
and
$$p_t=w_t \rho.$$
For the above linear forms of the EoS, the additional constraints are given as follows:
$$1>w_r>0,$$
and
$$1>w_t>0.$$
The graphical analysis of two parameters ($w_t$ and $w_r$) of the EoS is presented in Figs. (\ref{wr})-(\ref{wt}). This analysis shows that the linear form of the EoS we consider is in good agreement with the EoS of the ordinary matter distribution \cite{vikman2005can}. It is a crucial aspect that the parameters of the considered EoS have radial dependence only and do not depend on a constant quantity. These parameters are within bounds, and we can safely say that the matter inside the compact stars under investigation is ordinary charged matter.

\subsection{Anisotropic Measurement}
\begin{figure}[h!]
\centering
\epsfig{file=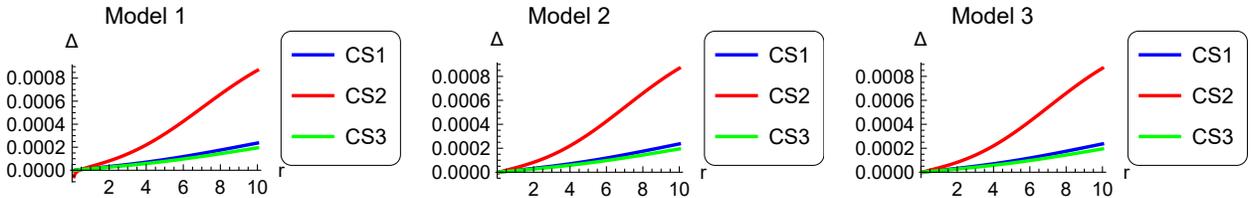,width=1\linewidth}
\caption{The change in anisotropy with respect to $r$}\label{ani}
\end{figure}

Anisotropy is a term that is commonly used in literature to describe the property of astronomical objects that exhibit different physical characteristics or behaviors when measured or observed from different directions or orientations. We take the following definition of anisotropy:
\begin{equation}\label{anisotropy}
\Delta  = \frac{2}{r}({p_t} - {p_r})
\end{equation}
For our study, we plotted the values of anisotropy obtained using Eq. (\ref{anisotropy}) with respect to the radial coordinate ($r$). The Fig. (\ref{ani}) shows the plots of anisotropy for considered compact stars in the framework of our suggested gravitational models. Using graphical analysis, we measure that in the case $p_t > p_r$, anisotropy is positive ($\Delta>0$), while in the case $p_t< p_r$, it is negative $\Delta<0$. The positive anisotropy implies that the direction of measurements is outward, while the negative anisotropy shows an inward direction. This behavior of anisotropy establishes that an anisotropic force can produce massive structures.

\subsection{Effects of Electric Field and Charge}
\begin{figure}[h!]
\centering
\epsfig{file=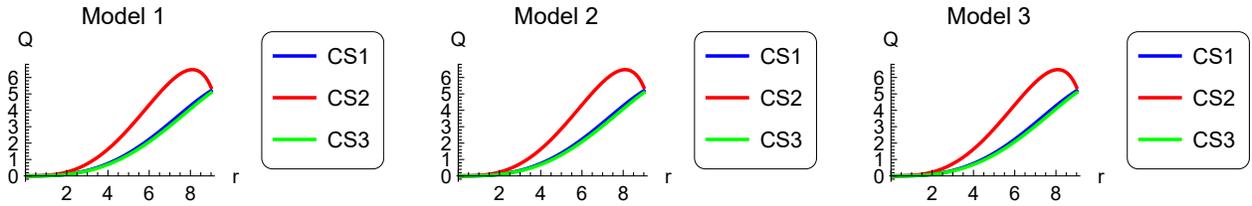,width=1\linewidth}
\caption{The effects of electric charge ($Q$) with changing radial coordinate}\label{charg}
\end{figure}

\begin{figure}[h!]
\centering
\epsfig{file=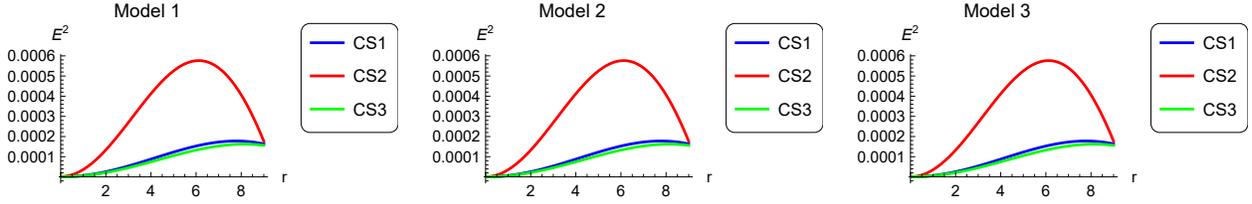,width=1\linewidth}
\caption{The effects of squared electric field ($E^2$) with changing radial coordinate}\label{electric}
\end{figure}

In this subsection of the article, we present some effects of the electric charge along with the electric field. We again use graphic description for analyzing these behaviors. The plots presented in Figs. (\ref{charg})-(\ref{electric}) show the impact of electric charge and field, respectively, for considered compact stars in the domain of our suggested gravitational models. Some very relevant and interesting effects of the electromagnetic field on the anisotropic collapse of compact stars have been investigated in Ref. \cite{Banerjee:2018yaa}.

\subsection{Analysis of Mass-Radius Relation, Compactness, and Redshift}
\begin{figure}[h!]
\centering
\epsfig{file=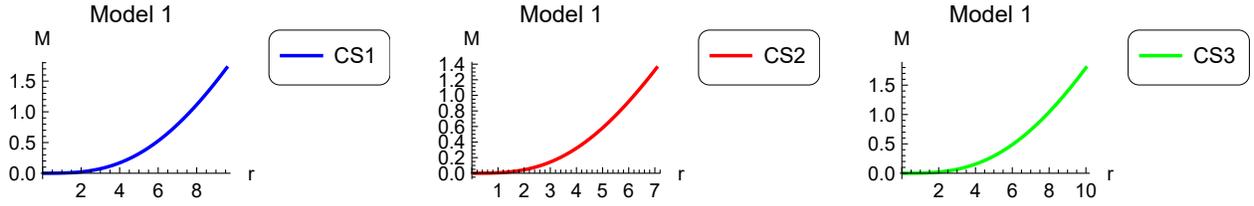,width=1\linewidth}
\caption{Mass function variations}\label{massvr}
\end{figure}

\begin{figure}[h!]
\centering
\epsfig{file=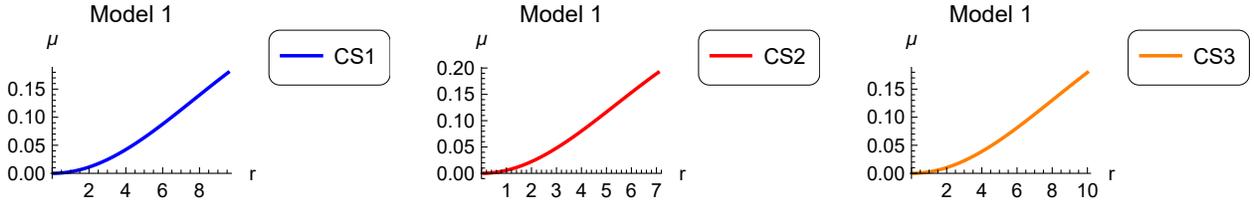,width=1\linewidth}
\caption{Compactness variations}\label{compact}
\end{figure}

\begin{figure}[h!]
\centering
\epsfig{file=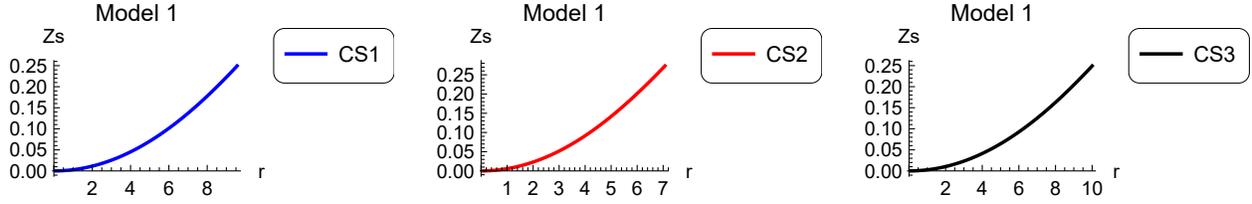,width=1\linewidth}
\caption{Redshift variations}\label{redshift}
\end{figure}

\begin{figure}[h!]
\centering
\epsfig{file=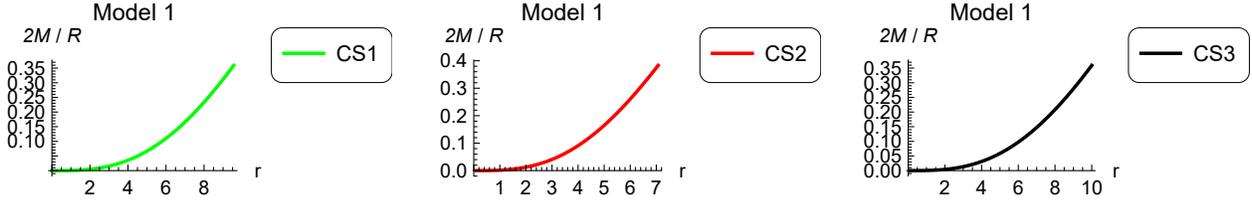,width=1\linewidth}
\caption{Buchdahl's limits for Model-I}\label{blimit}
\end{figure}

It is possible to express the mass of compact stars that carry an electric charge as a function of their radius. The mass-radius relation is a well-established concept that holds significant importance in comprehending the characteristics of compact stars. The EoS, which describes the interdependence between pressure, density, and temperature within the star, determines the mass-radius relation of a star. Understanding the correlation between mass and radius, as well as the EoS, is important in analyzing the characteristics of compact objects. Mathematically, it can be expressed as follows:
\begin{equation}\label{mrr}
m(r)=\int_{0}^{r} 4\pi \acute{r}^2\rho d\acute{r}.
\end{equation}
The above equation (\ref{mrr}) satisfy the conditions: for $r=0$, the $m(r)=0$ as $m(r=0)=0$ and for $r=R$, the $m(r)=M$ as $m(r=R)=M$. Figure (\ref{massvr}) illustrates the range of masses exhibited by charged compact stars. It is observed in Fig. (\ref{massvr}) that the mass distribution at the core is uniform, as it follows a direct proportionality with respect to the radial distance. Specifically, the mass function $m(r)$ approaches zero ($m(r)\rightarrow 0$) as the radial distance $r$ approaches zero ($r\rightarrow 0$), while the maximum value of this function is achieved at $r = R$.
Additionally, the compactness is mathematically expressed as follows:
\begin{equation}\label{cmpt}
\mu(r)=\frac{1}{r}\int_{0}^{r}4\pi \acute{r}^2 \rho d\acute{r}.
\end{equation}
The behavior of compactness as a function of radial distance is graphically demonstrated in Fig. (\ref{compact}).
Following the above two important concepts, the mathematical expression of redshift ($Z_s$) is given as follows:
\begin{equation}\label{reds}
  Z_s=\left(1-2\mu \right)^{-\frac{1}{2}}-1.
\end{equation}
The redshift is also plotted and corresponding variations are graphically demonstrated is Fig. (\ref{redshift}).
Extending our analysis, we have additionally checked the important Buchdahl's limit for Model-I; the plot for this limit is presented in Fig. (\ref{blimit}).

\section{Summary}
In this section, we present some key points and conclusive findings related to the current study. Mainly, we considered three compact stars, namely: Vela X-1, SAXJ 1808.4-3658, and 4U1820-30, for examining some important physical properties. Then we opted for the $f(R,G)$-MGT framework and suggested three distinct, feasible gravitational models to study the aforesaid physical properties of compact stars. Primarily, we assumed an anisotropic interior configuration for the considered compact objects. We then used matching techniques to match the interior geometry with the Reissner-Nordstrom exterior geometry. Through matching, we found expressions for arbitrary constants $A$, $B$, and $C$. Then the numerical values of these constants were computed using the approximate masses and radii of the considered compact stars. These values were further used to discuss the structure of compact stars. The following were concluded:
\begin{itemize}
  \item In the case of ordinary matter distribution, the parameters of the linear form of the EoS were found to be compatible for three compact stars under investigation in the domain of $f(R,G)$-gravity. The density, radial pressure, and transverse pressure functions were found to be maximal at the cores of stars while decreasing with the increase in radial distance. The positive components were discussed using the values of these functions. It was established that at the interior of compact stars, these components were finite. As a result, it was found that all the compact stars under investigation were singularity-free.
  \item Further, it was found that, when the transverse pressure is greater than the radial pressure, the anisotropy becomes positive, meaning that it is outwardly directed. Similarly, when the transverse pressure is less than the radial pressure,  the anisotropy becomes negative, meaning that it is inwardly directed.  Corresponding plots were also presented to establish this behavior and its results.
  \item Then the role of the TOV equation and the important ECs were investigated for the compact stars under investigation in the domain of suggested gravity models. The results of both TOV and ECs were presented using  graphical analysis.  In addition to this, we also presented plots for the stability condition ($|v_{st}^2-v_{sr}^2|<1$), which confirmed the stability and, hence, confirmed our models to be viable.
    \end{itemize}
It is interesting to extend this research by considering different possibilities; for instance, the behavior of these compact objects can be examined for a negative model parameter and in the domain of alternative MGTs, such as $f(R)$. The results and findings presented in the current study reasonably agree with the previous research.
\vspace{0.5cm}

\bibliographystyle{unsrt}
\bibliography{p7bib}
\end{document}